\documentclass[12pt,preprint]{aastex}

\newcommand{\etal}{$et~al.$}

\shorttitle{Organic Molecules in IRAS 16293-2422}
\shortauthors{Kuan, Huang, Charnley, Hirano, Takakuwa, et al.}

\begin{document}


\title{Organic Molecules in Low-Mass Protostellar Hot Cores:\\
	Submillimeter Imaging of IRAS 16293-2422}


\author{Yi-Jehng Kuan\altaffilmark{1,2}, Hui-Chun Huang\altaffilmark{1},
   Steven B. Charnley\altaffilmark{3}, \\
   Naomi Hirano\altaffilmark{2}, Shigehisa Takakuwa\altaffilmark{4},
   David J. Wilner\altaffilmark{5}, Sheng-Yuan Liu\altaffilmark{2}, \\
   Nagayoshi Ohashi\altaffilmark{2}, Tyler L. Bourke\altaffilmark{4},
   Chunhua Qi\altaffilmark{5}, Qizhou Zhang\altaffilmark{5}}




\altaffiltext{1}{Department of Earth Sciences, National Taiwan Normal
	University, 88 Sec.4 Ting-Chou Rd., Taipei 116, Taiwan,
	Republic of China; kuan, hspring@sgrb2.geos.ntnu.edu.tw}
\altaffiltext{2}{Academia Sinica Institute of Astronomy \& Astrophysics,
	P. O. Box 23-141, Taipei 106, Taiwan, ROC; hirano, ohashi, 
	syliu@asiaa.sinica.edu.tw}
\altaffiltext{3}{Space Science Division, MS 245-3, NASA Ames Research Center,
	Moffett Field, CA 94035; charnley@dusty.arc.nasa.gov}
\altaffiltext{4}{Harvard-Smithsonian Center for Astrophysics, Submillimeter
	Array Project, 645 N. A'ohoku Place, Hilo, HI 96721;
	stakakuw@sma.hawaii.edu}
\altaffiltext{5}{Harvard-Smithsonian Center for Astrophysics, 60 Garden
	Street, Cambridge, MA 02138; dwilner, tbourke, cqi, 
	qzhang@cfa.harvard.edu}


\begin{abstract}
Arcsecond-resolution spectral observations toward the protobinary system
IRAS 16293-2422 at 344 and 354 GHz were conducted using the Submillimeter
Array. Complex organic molecules such as CH$_3$OH and HCOOCH$_3$ were
detected.
Together with the rich organic inventory revealed, it clearly indicates
the existence of two, rather than one, compact {\it hot molecular cores}
($\lesssim$ 400 AU in radius) associated with each of the protobinary
components identified by their dust continuum emission in the inner
star-forming core.

\end{abstract}




\section{Introduction}

The low-mass protostellar source IRAS 16293-2422 (hereafter I16293) is
located in the $\rho$~Ophiuchus cloud complex at a distance 160 pc from
the Sun.  High angular-resolution observations revealed that I16293 is
also a protobinary system of two components with a projected separation
of $\sim$5.2\arcsec~($\sim$840 AU), and 3-mm continuum observations 
\citep{loo00} suggest that the northwest component (0.61 M$_{\sun}$,
hereafter I16293B) is slightly more massive than the southeast component
(0.49 M$_{\sun}$, hereafter I16293A).
%
%
Single-dish submillimeter line-surveys led \citet{van95}
to conclude that, within a 20\arcsec~region, I16293 consists of a cold
(T$_{\rm kin} \simeq$ 10$-$20 K) outer molecular envelope, a warmer
(T$_{\rm kin} \simeq$ 40 K) circumbinary envelope of
10\arcsec$-$15\arcsec~in size, and a hot (T$_{\rm kin} \gtrsim$ 80 K)
region of dense gas only 3\arcsec$-$10\arcsec~in size. This innermost gas
was found to be rich in organic molecules and it was suggested that their
presence was due to outflows \citep{hir01} interacting with the circumbinary
envelope. However, \citet{cec99,cea00}
showed that the emitting region is warm (100 K), dense $n_{\rm H_2}
\gtrsim 10^7~{\rm cm}^{-3}$, and very compact ($\sim$2\arcsec$-$3\arcsec).
This led to the suggestion that I16293 contained a {\it hot molecular
core} about 150 AU in size where high dust temperatures (T$_{\rm dust}
\gtrsim$ 100 K) can facilitate the evaporation of icy grain mantles
\citep{cea00}.  This conclusion is supported by recent observations which
show that I16293 {\it inner} hot core has a molecular inventory rich in
complex molecules, similar to the hot molecular cores (HMCs) associated 
with massive protostars \citep{caz03}.


Nevertheless, several questions remain that cannot easily be answered by
single-dish observations. One major concern is that its inferred size of
$\sim$150 AU ($\lesssim$~1\arcsec) \citep{cea00,sch02} is much smaller
than the binary separation in I16293, and comparable to the dimensions of
protostellar disks.  Furthermore, it is unknown whether high resolution
observations of low-mass hot cores will exhibit clear chemical 
differentiation on small spatial scales, similar to that seen in massive
HMCs \citep{wyr97}.


Submillimeter Array\footnote{The Submillimeter Array is a joint project
between the Smithsonian Astrophysical Observatory and the Academia Sinica
Institute of Astronomy and Astrophysics, and is funded by the Smithsonian
Institution and the Academia Sinica} observations with arcsecond resolution
(1\arcsec=160 AU) were carried out on 2003 March 14 (compact
configuration) and July 12 (extended configuration) with 5 antennae.
The phase tracking center of the observations was $\alpha$(J2000) =
16$^h32^m22^s$.91, $\delta$(J2000) = -24\arcdeg28\arcmin35\arcsec.52.
The digital correlator at the time was configured with eight overlapping
segments ("chunks") each of 104 MHz bandwidth and 128 channels, except
one chunk which had 512 channels, giving a frequency resolution of 0.812
and 0.203 MHz, respectively. The total bandwidth of each sideband was
$\sim$670 MHz covering two frequency ranges: 343.555$-$344.225 GHz (LSB)
and 354.211$-$354.881 GHz (USB). Quasars 1743-038 and NRAO530 were
observed for phase and amplitude calibration. The flux scale is estimated
to be accurate to 25\%. The synthesized beam sizes are 
$\sim$1\arcsec.3$\times$2\arcsec.7 at 344 GHz and
$\sim$1\arcsec.1$\times$2\arcsec.5 at 354 GHz with natural weighting;
these beams correspond to a linear scale of $\sim$200$\times$400 AU at a
distance of 160 pc.
The data were calibrated using the MIR software, and imaging was done
with MIRIAD software.

\section{Results and Discussions}

To make positive identifications of the spectral features detected, in
addition to the basic requirement of frequency coincidence, an iterative
process with the following stringent selection criteria was applied
to each candidate species:
1). Expected line intensity - unfavorably weak transitions were rejected.
2). Looking for transitions with similar or stronger line intensities in
    the bands - if missing, the particular candidate line was excluded.
3). Looking for slightly less favored transitions of the same molecule
for a possible sign of low-level ($<$ 3-$\sigma$) "detections" if no other
stronger transitions predicted were available in the bands. This would
be an indirect verification of the particular line identified.
4). Literature check against lists of known molecules in I16293 in the
    literature - not mandatory for a detection as SMA is sensitive to dense
    hot cores.
5). Making the requirement that the molecular candidate has to be a known
    hot core molecule.

Table~\ref{tbl-1} lists some of the representative molecular transitions
detected.
The LSR velocity of each molecular line measured is also listed
in Table~\ref{tbl-1}. Single-dish observations reveal that the LSR
velocity of I16293 is centered at 3.9 to 4.2 km s$^{-1}$ and is often
found within the range of 3.3$-$4.7 km s$^{-1}$.
Asymmetric double-peak line profiles with the blueshifted peak at
2.5$-$3.5 km s$^{-1}$ and the red one at 5.0$-$6.0 km s$^{-1}$ are
readily seen in single-dish observations.

\begin{figure}
%
\hspace{+5mm}
\includegraphics[width=2.5in]{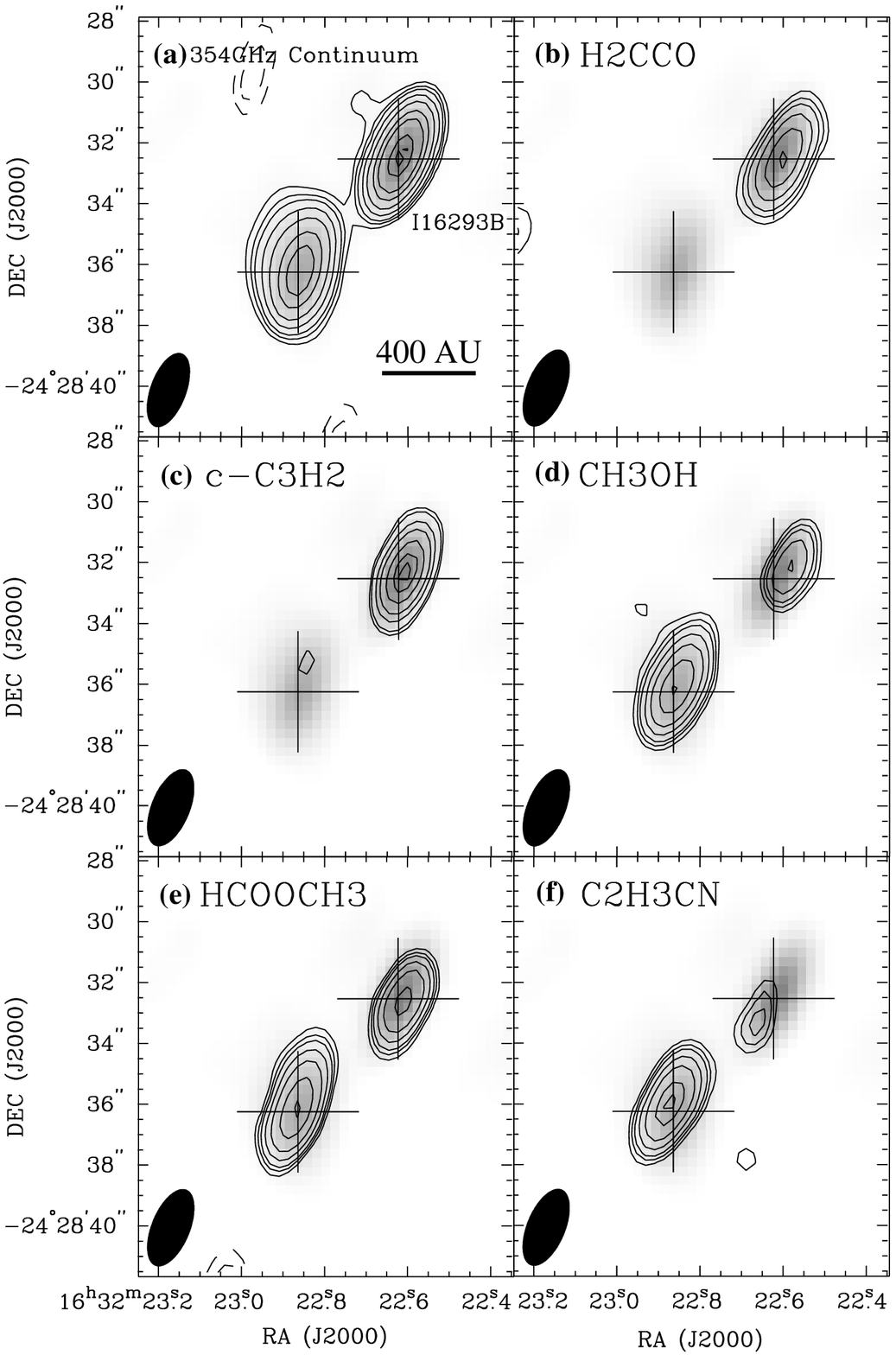}
%
\caption{Spectral images of large organic molecules toward I16293.
(a) Continuum at 354 GHz; (b) spectral emission of CH$_2$CO;
(c) $c$-C$_3$H$_2$ emission; (d) CH$_3$OH; (e) HCOOCH$_3$ at 344029 MHz;
and (f) CH$_2$CHCN.
Crosses mark the positions of I16293A and I16293B hot cores. The angular
size for a linear scale of 400 AU is shown in (a). The grey scale denotes
the 354 GHz continuum. Continuum emission was imaged with the wide
continuum channel at 354 GHz, at a noise level of 0.11 Jy beam$^{-1}$,
and the estimated line contamination is $<$ 3\% in I16293A and $<$ 1\% in
I16293B. The continuum flux density of I16293A is 4.97$\pm$0.49 Jy, which
is slightly lower than that of I16293B (5.14$\pm$0.61 Jy).  The dark
ellipse represents the HPBW of the synthesized beam.  Contours are shown
at 3-, 4-, 5-, 7-$\sigma$ levels in general, then at irregular intervals
up to the peak values; dashed lines indicate contours at negative 3- and
4-$\sigma$ levels.
\label{fig1}}

\end{figure}

\begin{figure}
\hspace{+4mm}
\includegraphics[width=3.0in]{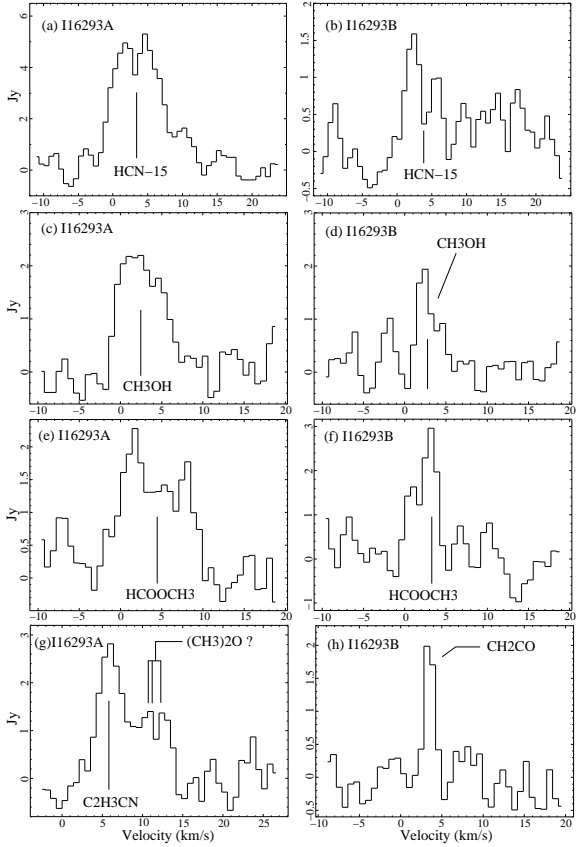}
\caption{Sample spectra of large organic molecules toward I16293.
The left column shows spectra taken at I16293A; the right column,
at I16293B. HC$^{15}$N spectra are shown in (a) and (b); CH$_3$OH,
in (c) and (d); and HCOOCH$_3$ (344029 MHz), in (e) and (f).
(g) gives C$_2$H$_3$CN line and (h), CH$_2$CO.
Tentatively detected (CH$_3$)$_2$O 17$_{2,16}$-16$_{1,15}$
EA, AE, EE \& AA transitions at 343753.3, 343754.2 and 343755.1 MHz
can also be seen in (g). The weaker spectral appearance of the 
tentative (CH$_3$)$_2$O line is because spectrum (g) is taken at the
peak position of the integrated CH$_2$CHCN line. All spectra were
Hanning smoothed for better S/N ratios.\label{fig2}}

\end{figure}

On the other hand, an interferometer like the SMA can easily filter out
the diffuse ambient cloud, such as the warm gas envelope of I16293 at
V$_{\rm LSR}$ $\sim$3.9$-$4.2 km s$^{-1}$, and its small beam will tend
to pick up the fine kinematic details of dense hot regions. The SMA beam
is small enough to resolve the two protobinary components in I16293, thus
SMA is able to trace individual velocity component in the rotating disks
and/or hot cores.
As shown in Fig.~\ref{fig1}$a$, the two protobinary components are clearly
resolved in the continuum emission. Sample images of five organic molecules
are shown in Figs.~\ref{fig1}{\it b-f}.
Fig.~\ref{fig2} shows the sample spectra of four large organic molecules
plus HC$^{15}$N in I16293A and I16293B. By examining HC$^{15}$N emission
toward I16293A, two major velocity components, one at 1.7 km s$^{-1}$ and
the other at 4.7 km s$^{-1}$ were seen (Fig. 2$a$); in I16293B, one 
component at 2.5 and the other at 6.0 km s$^{-1}$ were also exhibited
(Fig. 2$b$).
%
%
Furthermore, the small beam of an array
suffers less beam dilution, thus is sensitive to high-velocity components
within compact sources. In the case of HCN 4$-$3 and HC$^{15}$N, for
example, linewidths $\gtrsim$8 km s$^{-1}$ are seen toward I16293A.
The HCOOCH$_3$ line in I16293A is also fairly wide, with a linewidth of 
$\sim$8 km s$^{-1}$ (Fig.~\ref{fig2}$e$). The large linewidth (5$-$8 
km s$^{-1}$) is common for simple molecules like HCN, or strong molecular
emission like HC$_3$N and CH$_3$OH. For weaker emission, typical linewidths
are of 2$-$4 km s$^{-1}$ in I16293A; however, toward I16293B, the 
linewidths of most molecular lines are notably narrower, i.e., between 2
and 4 km s$^{-1}$.

Because of array characteristics and sensitivity, it is not unusual that
an array can preferentially pick up only the stronger emission peak of a
double-peak profile as seen by a single-dish.
The CH$_3$OH line shows V$_{\rm LSR}$ $\sim$ 2.5
km s$^{-1}$ toward I16293A, with two velocity components visible near
4.6 and 1.8 km s$^{-1}$ (Fig.~\ref{fig2}$c$). The kinematic structure
traced is thus similar to what seen by HC$^{15}$N; a similar line profile
of $^{13}$CH$_3$OH was also observed toward I16293A. Toward I16293B,
the CH$_3$OH emission is weaker and is mainly from the blueshifted
component near 2.5 km s$^{-1}$ (Fig.~\ref{fig2}$d$), which again 
agrees well with what was seen by HC$^{15}$N. Since the optically thin
HC$^{15}$N emission basically traced the densest regions such as the
disks, it suggests methanol may also reside in the same dense regions.
However, it is also plausible that the two velocity components
seen in the methanol line profile are due to self absorption at
$\sim$3.9 km s$^{-1}$.
Toward I16293B, the LSR velocities of HCOOCH$_3$ lines were found 
consistent with each other at $\sim$3.9 km s$^{-1}$, within the velocity
uncertainty of $\sim$0.7 km s$^{-1}$. However, a non-trivial velocity
scattering with V$_{\rm LSR}$ between 3.9 and 6.0 km s$^{-1}$ was 
observed toward I16293A. A close examination of the integrated intensity
maps of HCOOCH$_3$ show their peak emission positions are situated at
$\sim$0\arcsec.5 east (V$_{\rm LSR}$ = 3.9 km s$^{-1}$) and south 
(V$_{\rm LSR}$ = 5.9 and 6.0 km s$^{-1}$) of the I16293A center position
where V$_{\rm LSR}$ $\sim$ 4.6 km s$^{-1}$. Again, such a velocity field
variation traces the general velocity gradient seen by the HC$^{15}$N
%
%
emission nicely.


All spectral emission appears to originate from two compact
regions of a beam-convolved size of $\sim$200$\times$400 AU in radius,
concentrated toward the two protobinary components I16293A and I16293B.
An exception is the 4$-$3 HCN transition which
shows complicated kinematic structure (Takakuwa \etal~2004, in preparation).

\begin{figure}
\hspace{+4mm}
\includegraphics[angle=-90,width=1.8in]{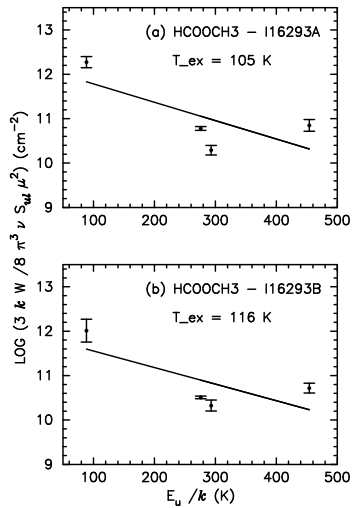}
\caption{The rotation diagrams of HCOOCH$_3$. Note that the 344170 MHz
doublet (Eu = 454 K) appears to be situated at a location slightly higher
in the rotation-diagram fits. It might imply this doublet is just in the
transitional regime toward a higher excitation,
or it might indicate simply this is a misidentified line.
Excluding this doublet would result in a lower excitation temperature,
which agrees better with the value ($\sim$60 K) of \citet{caz03}.
\label{fig3}}

\end{figure}

The beam-averaged line-of-sight column densities and fractional
abundances of all observed molecules are listed in Table~\ref{tbl-2},
where a gas column density $N_{\rm H_2} = 1.6 \times 10^{24}$ cm$^{-2}$
\citep{sch02} was adopted for both protostellar cores.  Column densities
were measured at the peak emission positions of each integrated spectral
line and were derived assuming thermalized level populations and optically
thin lines.
At these high
densities, gas and dust are expected to be thermally well-coupled and so
an excitation temperature T$_{\rm ex}$ $\simeq$~T$_{\rm dust}$ = 100 K was
adopted, except for molecular transitions with energy levels $>$ 500
cm$^{-1}$ ($\gtrsim$700 K) where $T_{\rm ex}$ = 300 K was used 
\citep{sch02}.
For HCOOCH$_3$, with multi-line detections, the column densities were
derived from rotation diagrams (Fig.~\ref{fig3}); we found T$_{\rm ex}$ =
105 K for I16293A, and 116 K for I16293B. The assumption of T$_{\rm ex}$
= 100 K applied for transitions of lower energy levels is hence warranted. 

The derived fractional abundances and relative column densities are in good
agreement with those derived from submillimeter observations of the Orion KL
and Sgr B2 massive HMCs \citep{sut91,sut95}.
When molecular emission is mainly from the compact cores, abundances
derived using the smaller SMA beam can be much higher (e.g. $c$-C$_3$H$_2$)
than found in previous single-dish observations \citep{van95,caz03}.
%
%
This explains why CH$_3$OH abundances derived from modeling higher energy
%
%
transitions \citep{sch02} in the {\it inner} I16293 core show good
agreement with our values, and suggests that the lower array abundances
for HCN and HCOOCH$_3$ is indicative of less compact emission where the
extended component is filtered out by the interferometer.

The high abundances of organic molecules, particularly methanol, indicates
that icy mantles have recently been evaporated \citep{cha92}.
%
%
For many molecules detected in both sources, the measured abundances only
differ by factors of about 2$-$3; as expected if both cores collapsed from
the same cold core material.  Nevertheless, the apparent absence of
$c$-C$_3$H$_2$ and CH$_2$CO in I16293A, if confirmed, raises the prospect
that future observations could discover pronounced chemical differentiation
between the two cores.

We have resolved the hot cores surrounding both protostellar sources
in I16293 and measured the chemical composition in each.  These
preliminary observations demonstrate that the SMA will be an important
tool to further explore the connection between the volatile organic
chemical composition of the hot cores associated with massive protostars,
those of Solar mass, and the composition of comets.

\acknowledgments

This work was supported by NSC 92-2112-M-003-006 grant (Y.-J.K.) and
by NASA's Exobiology and Origins of Solar Systems Programs through NASA
Ames Cooperative Agreement No. NCC2-1412 (S.B.C.).

\clearpage





\clearpage

\clearpage

\clearpage

\begin{deluxetable}{llcrrcrc}
\tabletypesize{\scriptsize}
\tablecaption{Sample molecules detected toward IRAS 16293-2422 protostellar
cores.\label{tbl-1}}
\tablewidth{0pt}
\tablehead{
\colhead{Molecule} & \colhead{Transition} & \colhead{Frequency} &
  \colhead{E$_{low}$} & \colhead{I$_{\nu,{\rm I16293A}}$\tablenotemark{a}} &
  \colhead{$V_{\rm LSR, A}$\tablenotemark{b}} &
  \colhead{I$_{\nu,{\rm I16293B}}$\tablenotemark{a}} &
  \colhead{$V_{\rm LSR, B}$\tablenotemark{b}} \\
 & & \colhead{(MHz)} & \colhead{(cm$^{-1}$)} & \colhead{(Jy bm$^{-1}$)} &
 \colhead{(km s$^{-1}$)} & \colhead{(Jy bm$^{-1}$)} & \colhead{(km s$^{-1}$)}
}
\startdata
HCN & 4-3 & 354505.5 & 17.7 & 10.88$\pm$0.44 & 4.6 & 9.92$\pm$0.44 & 3.6 \\
HC$^{15}$N & 4-3 & 344200.3 & 17.2 & 5.26$\pm$0.24 & 3.9 &
	1.72$\pm$0.24 & 3.9 \\
$c$-C$_3$H$_2$ & 23$_{13,10}$-23$_{12,11}$ & 343804.9 & 548.9 &
	2-$\sigma$\tablenotemark{c} & \nodata & 3.48$\pm$0.32 & 4.6 \\
CH$_2$CO & 17$_{2,15}$-16$_{2,14}$ & 343693.9 & 127.9 &
        2-$\sigma$ & \nodata & 1.98$\pm$0.28 & 3.5 \\
HC$_3$N & 39-38 & 354697.4 & 224.8 & 3.30$\pm$0.20 & 4.6 &
	1.54$\pm$0.20 & 3.9 \\
CH$_3$OH & 18$_{2,16}$-17$_{3,14}$ E & 344109.1 & 280.0 &
	2.20$\pm$0.15 & 2.5 & 1.94$\pm$0.15 & 2.5 \\
$^{13}$CH$_3$OH & 4$_{1,3}$-3$_{0,3}$ A & 354445.9 & 18.6 &
	2.27$\pm$0.20\tablenotemark{d} & 2.5 & 2-$\sigma$ & \nodata \\
C$_2$H$_3$CN ($v_{15}$=1) & 36$_{4,32}$-35$_{4,31}$ & 343761.8 & 564.9 &
	2.82$\pm$0.26 & 5.3 & 0.94$\pm$0.26 & 5.3 \\
HCOOCH$_3$ & 32$_{*,32}$-31$_{*,31}$ A & 344029.6\tablenotemark{e} & 180.4 &
		2.28$\pm$0.16 & 4.6 & 2.96$\pm$0.16 & 3.2 \\
   & 32$_{2,31}$-31$_{1,31}$ E & 344029.8\tablenotemark{f} & 180.4 & & & & \\
         & 28$_{18,10}$-27$_{18,9}$ A & 344170.9 & 304.2 &
		1.09$\pm$0.20 & 6.0 & 1.04$\pm$0.20 & 3.9 \\
         & 28$_{18,11}$-27$_{18,10}$ A & 344170.9 & 304.2 & & & & \\
         & 33$_{*,33}$-32$_{*,32}$ A & 354608.0\tablenotemark{e} & 191.9 &
		1.71$\pm$0.20 & 5.9 & 2.75$\pm$0.20 & 3.9 \\
   & 33$_{2,32}$-32$_{1,32}$ E & 354608.4\tablenotemark{g} & 191.9 & & & & \\
        & 12$_{8,5}$-11$_{7,5}$ E   & 354742.4 & 49.7  & 1.85$\pm$0.19 & 3.9 &
		0.98$\pm$0.19 & 3.9 \\
\enddata


\tablenotetext{a}{~The line intensity measured in Hanning-smoothed spectrum
  at the peak-emission position of integrated intensity map. This value is
  generally lower than the actual peak intensity shown in channel maps due
  to Hanning smooth.}
\tablenotetext{b}{~The uncertainty of the LSR velocity determined is
  $\sim$0.7 km s$^{-1}$.}
\tablenotetext{c}{~Only a $\sim$2-$\sigma$ ($\lesssim$ 0.6 Jy beam$^{-1}$)
  detection in channel maps.}
\tablenotetext{d}{~Partially blended with weak HCOOH $b$-type transition
		17$_{0,17}$-16$_{1,16}$ at 354448.3 MHz.}
\tablenotetext{e}{~* = 0, 1.}
\tablenotetext{f}{~Together with transitions 32$_{1,32}$-31$_{1,31}$ E,
		32$_{2,31}$-31$_{2,30}$ E and 32$_{1,32}$-31$_{2,30}$ E.}
\tablenotetext{g}{~Together with transitions 33$_{1,33}$-32$_{1,32}$ E,
		33$_{2,32}$-32$_{2,31}$ E and 33$_{1,33}$-32$_{2,31}$ E.}

%
%

\end{deluxetable}

\clearpage

\begin{deluxetable}{lccccccccc}
\tabletypesize{\scriptsize}
\tablecaption{Molecular column densities and fractional abundances
	toward IRAS 16293-2422.\label{tbl-2}}
\tablewidth{0pt}
\tablehead{
 & & \colhead{I16293A} & & & & \colhead{I16293B} & & \colhead{HMC} &
     \colhead{IRAS 16293} \\
\cline{2-4}  \cline{6-8}
\colhead{Molecule} & \colhead{$\int$ I$_{\nu} dV$\tablenotemark{a}} &
     \colhead{$N$\tablenotemark{b}} & \colhead{$X$} & &
     \colhead{$\int$ I$_{\nu} dV$\tablenotemark{a}} &
     \colhead{$N$\tablenotemark{b}} & \colhead{$X$} & \colhead{$X$} &
     \colhead{$X$} \\
   & & \colhead{(cm$^{-2}$)} & \colhead{($N/N_{{\rm H}_2}$)} & &
   & \colhead{(cm$^{-2}$)} & \colhead{($N/N_{{\rm H}_2}$)} &
     \colhead{($N/N_{{\rm H}_2}$)} & \colhead{($N/N_{{\rm H}_2}$)}
}

\startdata
HCN  & 91.80 & 3.1(+14) & 2.0(-10) & & 49.03 & 1.7(+14) & 1.0(-10) &
   3.2(-9)\tablenotemark{c} & 1.9(-9)\tablenotemark{d} \\
HC$^{15}$N  &  42.86 & 1.2(+14) & 7.4(-11) & & 6.97 & 1.9(+13) & 1.2(-11) &
   \nodata & 7.0(-12)\tablenotemark{d} \\
$c$-C$_3$H$_2$  &   \nodata  &  \nodata  &  \nodata  & & 10.03 &
   7.2(+15)\tablenotemark{e} & 4.5(-9) & 6.3(-11)\tablenotemark{c} &
   3.5(-11)\tablenotemark{d} \\
CH$_2$CO  & \nodata  &  \nodata  &  \nodata  & & 3.22 & 1.9(+15) & 1.2(-9) &
   3.(-10)\tablenotemark{f} &
   1.8(-10)\tablenotemark{d} , 5.0(-8)\tablenotemark{g} \\
HC$_3$N  &  16.73 & 6.7(+14) & 4.2(-10) & & 3.34 & 1.3(+14) & 8.4(-11) &
   1.8(-9)\tablenotemark{f} &
   2.5(-11)\tablenotemark{d} , 1.0(-9)\tablenotemark{g} \\
CH$_3$OH & 13.62 & 1.1(+18) & 6.8(-7) & & 6.20 & 5.0(+17) & 3.1(-7) &
   1.4(-7)\tablenotemark{f} &
   4.4(-9)\tablenotemark{d} , 3.0(-7)\tablenotemark{g}\\
$^{13}$CH$_3$OH &  17.03 & 8.1(+16) & 5.0(-8)  & &
   \nodata  &  \nodata  &  \nodata  &  \nodata  & \nodata    \\
CH$_2$CHCN & 8.58 & 1.5(+16)\tablenotemark{e} & 9.4(-9)  & & 2.35  &
   4.1(+15)\tablenotemark{e} & 2.6(-9) & 1.5(-9)\tablenotemark{f} & \nodata \\
HCOOCH$_3$ & ---\tablenotemark{h} & 6.8(+15) & 4.3(-9) & &
   ---\tablenotemark{h} & 4.2(+15) & 2.6(-9) &
   1.4(-8)\tablenotemark{f} & 2(-7)\tablenotemark{i} \\
\enddata

\tablenotetext{a}{~In Jy bm$^{-1}$~km s$^{-1}$.}
\tablenotetext{b}{~T$_{\rm ex}$ = 100 K was assumed for all lines except
    where noted. For HCOOCH$_3$, column densities were derived from actual
    rotation-diagram fits.}
\tablenotetext{c}{~Sgr B2(M) hot core; from \citet{sut91}.}
\tablenotetext{d}{~Single-dish observations with a $\sim20\arcsec$~beam;
    from \citet{van95}.}
\tablenotetext{e}{~For transitions with energy levels $>$ 500 cm$^{-1}$,
    T$_{\rm ex}$ = 300 K is adopted.}
\tablenotetext{f}{~Orion KL hot core; from \citet{sut95}.}
\tablenotetext{g}{~IRAS 16293 hot core; from models of \citet{sch02}.}
\tablenotetext{h}{~In I16293B, the integrated intensities of HCOOCH$_3$
   lines are 8.71, 3.14, 5.55 and 2.31 Jy bm$^{-1}$~km s$^{-1}$ with
   increasing frequency, and in I16293A, 16.22, 4.25, 5.12 and 4.22 
   Jy bm$^{-1}$~km s$^{-1}$.}
\tablenotetext{i}{~Single-dish observations with $\sim10\arcsec-30\arcsec$
   					     beam; from \citet{caz03}.}

\end{deluxetable}

\clearpage

\end{document}